# COMPOSITION OF GRAY ISOMETRIES

## SIERRA MARIE M. LAURESTA and Virgilio P. Sison

Institute of Mathematical Sciences and Physics
University of the Philippines Los Baños
College, Laguna 4031
E-mail: vpsison@up.edu.ph, vpsison@uplb.edu.ph

**Abstract**

In coding theory, Gray isometries are usually defined as mappings between finite Frobenius rings, which include the ring $\mathbb{Z}_m$ of integers modulo *m* and the finite fields. In this paper, we derive an isometric mapping from $\mathbb{Z}_8$ to $\mathbb{Z}_4^2$ from the composition of the Gray isometries on $\mathbb{Z}_8$ and on $\mathbb{Z}_4^2$. The image under this composition of a $\mathbb{Z}_8$-linear block code of length *n* with homogeneous distance *d* is a (not necessarily linear) quaternary block code of length *2n* with Lee distance *d*.

**Introduction**

A **(block) code** of length *n* over a ring *R* is a non-empty set of *n*-tuples over *R* called **codewords**. It is said to be **linear** of **rate-*k/n*** if it is a (not necessarily free) submodule, and is completely determined by a $k \times n$ matrix *G* over *R*.

Codes over rings gained more attention when Hammons, Kumar, Calderbank, Sloane and Solé [1] discovered in 1994 that certain very good but peculiar nonlinear codes over the binary field $\mathbb{F}_2$ can be viewed as images of linear codes over the integer ring $\mathbb{Z}_4$ under the so-called Gray map $\phi$ from $\mathbb{Z}_4$ onto $\mathbb{F}_2^2$ defined by $0 \mapsto (0,0)$, $1 \mapsto (0,1)$, $2 \mapsto (1,1)$ and $3 \mapsto (1,0)$. The map $\phi$ is an isometry or is weight-preserving, that is, the Lee weight of an element of $\mathbb{Z}_4$ is equal to the Hamming weight of its image under $\phi$. The Hamming weight of a binary vector is the number of nonzero components in the vector.

Carlet [2] introduced a generalization of $\phi$ to the ring $\mathbb{Z}_{2^k}$ of integers modulo $2^k$. Let *k* be a positive integer, *u* an element of $\mathbb{Z}_{2^k}$ and $u = \sum_{i=1}^{k} 2^{i-1} u_i$ its 2-adic expansion, where $u_i \in \mathbb{F}_2$. The image of *u* by the generalized Gray map is the boolean function on $\mathbb{F}_2^{k-1}$ given by

$$(y_1, y_2, \ldots, y_{k-1}) \longrightarrow u_k + \sum_{1=1}^{k-1} u_i y_i.$$

We identify this boolean function with a binary word of length $2^{k-1}$ by simply listing its values. Thus the generalized Gray map is seen as a nonsurjective mapping from $\mathbb{Z}_{2^k}$ to $\mathbb{F}_2^{2^{k-1}}$, and its image is the Reed-Muller code of order 1, *RM(1, k-1)*. The generalized Gray map is naturally extended to the *n*-tuples.

When *k* = 2, *RM(1,1)* is the set of boolean functions on $\mathbb{F}_2$, and we obtain the usual Gray map $\phi$ from $\mathbb{Z}_4$ to $\mathbb{F}_2^2$. When *k* = 3, the generalized Gray map (which we denote by $\psi$) takes $\mathbb{Z}_8$ onto *RM(1,2)* which is the set of boolean functions on $\mathbb{F}_2^2$ that give all the binary words in $\mathbb{F}_2^4$ with even Hamming weight.

## Methodology

We extend the usual Gray isometry $\phi$ as a bijective mapping from $\mathbb{Z}_4^2$ onto $\mathbb{F}_2^4$. Table 1 shows the binary image of an element of $\mathbb{Z}_4^2$ under $\phi$. Clearly the Lee weight $w_L$ of an element of $\mathbb{Z}_4^2$ is equal to the Hamming weight $w_H$ of its binary image.

| $(x,y) \in \mathbb{Z}_4^2$ | $w_L(x,y)$ | $\phi(x,y) \in \mathbb{F}_2^4$ | $w_H(\phi(x,y))$ |
|---|---|---|---|
| (0,0) | 0 | (0,0,0,0) | 0 |
| (0,1) | 1 | (0,0,0,1) | 1 |
| (0,2) | 2 | (0,0,1,1) | 2 |
| (0,3) | 1 | (0,0,1,0) | 1 |
| (1,0) | 1 | (0,1,0,0) | 1 |
| (1,1) | 2 | (0,1,0,1) | 2 |
| (1,2) | 3 | (0,1,1,1) | 3 |
| (1,3) | 2 | (0,1,1,0) | 2 |
| (2,0) | 2 | (1,1,0,0) | 2 |
| (2,1) | 3 | (1,1,0,1) | 3 |
| (2,2) | 4 | (1,1,1,1) | 4 |
| (2,3) | 3 | (1,1,1,0) | 3 |
| (3,0) | 1 | (1,0,0,0) | 1 |
| (3,1) | 2 | (1,0,0,1) | 2 |
| (3,2) | 3 | (1,0,1,1) | 3 |
| (3,3) | 2 | (1,0,1,0) | 2 |

Table 1: The isometric Gray map $\phi$ on $\mathbb{Z}_4^2$

We restrict $\phi^{-1}$ as a mapping from *RM(1,2)* to $\mathbb{Z}_4^2$ as follows:

| $x \in RM(1,2)$ | $\phi^{-1}(x)$ |
|---|---|
| (0,0,0,0) | (0,0) |
| (0,1,0,1) | (1,1) |
| (0,0,1,1) | (0,2) |
| (0,1,1,0) | (1,3) |
| (1,1,1,1) | (2,2) |
| (1,0,1,0) | (3,3) |
| (1,1,0,0) | (2,0) |
| (1,0,0,1) | (3,1) |

Table 2: The map $\phi^{-1}$ on $RM(1,2)$

For $\mathbb{Z}_8$ we apply the following homogeneous weight [3] and extend it coordinatewise.

$$w_{hom}(x) = \begin{cases} 0 & if\ x = 0 \\ 4 & if\ x = 4 \\ 2 & else \end{cases}$$

Table 3 shows the image of an element of $\mathbb{Z}_8$ in *RM(1,2)* under the generalized Gray map $\psi$. If $u \in \mathbb{Z}_8$ has the 2-adic expansion $u = u_1 + 2u_2 + 4u_3$, then $\psi(u) = \psi(u_1 + 2u_2 + 4u_3) = (u_3, u_3 + u_1, u_3 + u_2, u_3 + u_1 + u_2)$. The mapping $\psi$ is weight preserving such that the homogeneous weight of an element of $\mathbb{Z}_8$ is equal to the Hamming weight of its image in *RM(1,2)*.

| $u \in \mathbb{Z}_8$ | $w_{hom}(u)$ | $\psi(u) \in RM(1,2)$ | $w_H(\psi(u))$ |
|---|---|---|---|
| 0 | 0 | (0,0,0,0) | 0 |
| 1 | 2 | (0,1,0,1) | 2 |
| 2 | 2 | (0,0,1,1) | 2 |
| 3 | 2 | (0,1,1,0) | 2 |
| 4 | 4 | (1,1,1,1) | 4 |
| 5 | 2 | (1,0,1,0) | 2 |
| 6 | 2 | (1,1,0,0) | 2 |
| 7 | 2 | (1,0,0,1) | 2 |

Table 3: The isometric Gray map $\psi$ on $\mathbb{Z}_8$

**Results and Discussion**

We take the composition $\phi^{-1}\psi: \mathbb{Z}_8 \rightarrow \mathbb{Z}_4^2$. Table 4 shows the quaternary image of an element of $\mathbb{Z}_8$ under $\phi^{-1}\psi$. If $u = u_1 + 2u_2 + 4u_3 \in \mathbb{Z}_8$, then $\phi^{-1}\psi(u) = \phi^{-1}\psi(u_1 + 2u_2 + 4u_3) = (u_1 + 2u_3, u_1 + 2u_3 + 2u_2)$.

| $u \in \mathbb{Z}_8$ | $w_{hom}(u)$ | $\phi^{-1}\psi(u)$ | $w_L(\phi^{-1}\psi(u))$ |
|---|---|---|---|
| 0 | 0 | (0,0) | 0 |
| 1 | 2 | (1,1) | 2 |
| 2 | 2 | (0,2) | 2 |
| 3 | 2 | (1,3) | 2 |
| 4 | 4 | (2,2) | 4 |
| 5 | 2 | (3,3) | 2 |
| 6 | 2 | (2,0) | 2 |
| 7 | 2 | (3,1) | 2 |

Table 4: The isometric map $\phi^{-1}\psi: \mathbb{Z}_8 \rightarrow \mathbb{Z}_4^2$

The mapping $\phi^{-1}\psi$ is weight preserving such that the homogeneous weight of an element of $\mathbb{Z}_8$ is equal to the Lee weight of its image in $\mathbb{Z}_4^2$. It is extended naturally to the *n*-tuples.

Let $C$ be a linear block code of length *n* over $\mathbb{Z}_8$ with minimum homogeneous distance *d*. The image of $C$ under $\phi^{-1}\psi$ is the set $\phi^{-1}\psi(C) = \{w \in \mathbb{Z}_4^{2n} \mid w = \phi^{-1}\psi(c) \text{ for } c \in C\}$.

*Proposition*. The set $\phi^{-1}\psi(C)$ has the following properties:

   i. $\phi^{-1}\psi(C)$ is a (not necessarily linear) block code of length *2n* over $\mathbb{Z}_4$.
   ii. The Lee distance of $\phi^{-1}\psi(C)$ is equal to *d*.
   iii. Every codeword of $\phi^{-1}\psi(C)$ has even Lee weight.

To illustrate non-linearity, consider the rate-2/3 linear block code over $\mathbb{Z}_8$ generated by the matrix

$$G = \begin{pmatrix} 1 & 2 & 7 \\ 0 & 2 & 4 \end{pmatrix}$$

This code has 32 codewords, minimum Hamming distance 1, and minimum homogeneous distance 4. The codewords (6,6,6) and (7,6,1) generated by the information words (6,1) and (7,4), respectively, have quaternary images (2,0,2,0,2,0) and (3,1,2,0,1,1) whose superimposition is not in the code. This example also shows that $\phi^{-1}\psi$ is not an additive homomorphism.

**Conclusion and Recommedation**

This paper offers a simple way to define isometric mappings from $\mathbb{Z}_{2^k}$ to $\mathbb{Z}_{2^r}$, in general for $r < k$, and to take the $\mathbb{Z}_{2^r}$-image code of a $\mathbb{Z}_{2^k}$-linear block code. Sufficient and necessary conditions for the linearity of the $\mathbb{Z}_{2^r}$-image can be determined. Extension of this construction to galois rings is inevitable.